\documentclass[runningheads]{llncs}

\usepackage[T1]{fontenc}
\usepackage{graphicx} 
\usepackage{float}
\usepackage{hyperref} 
\usepackage{tikz} 
\usetikzlibrary{arrows.meta, positioning, shapes.geometric, calc}
\usepackage{amsmath, amssymb, amsfonts} 
\usepackage{booktabs} 
\usepackage{textcomp}
\usepackage[sorting=none]{biblatex}
\usepackage{xcolor}
\usepackage{caption} 
\usepackage{algorithmic}
\usepackage{tabularx}
\usepackage{pgfplots} 
\pgfplotsset{compat=1.18}
\usepackage{orcidlink}
\begin{document}

\title{FLUID: Flow-Latent Unified Integration via Token Distillation for Expert Specialization in Multimodal Learning}
\author{
Van Duc Cuong\inst{1}\orcidID{0009-0008-6700-1477} \and
Ta Dinh Tam\inst{1}\orcidID{0009-0007-6567-4304} \and
Tran Duc Chinh\inst{1}\orcidID{0009-0002-8078-6519} \and
Nguyen Thi Hanh\inst{2, 3}\orcidID{0009-0006-7929-3329}
}
\institute{
Hanoi University of Science and Technology, Hanoi, Vietnam \and
Faculty of Interdisciplinary Digital Technology, PHENIKAA University, Hanoi, Vietnam; \and 
Corresponding author at PHENIKAA University, Hanoi, Vietnam\\
\email{\{cuong.vd220021, tam.td2400071, chinh.td224936\}@sis.hust.edu.vn, hanh.nguyenthi@phenikaa-uni.edu.vn}
}

\authorrunning{Cuong et al.}

\titlerunning{FLUID}


\maketitle
\begin{abstract}
Multimodal classification requires robust integration of visual and textual signals, yet common fusion strategies are brittle and vulnerable to modality-specific noise. In this paper, we present \textsc{FLUID}-Flow-Latent Unified Integration via Token Distillation for Expert Specialization, a principled token-level pipeline that improves cross-modal robustness and scalability. \textsc{FLUID} contributes three core elements: (1) \emph{Q-transforms}, learnable query tokens that distill and retain salient token-level features from modality-specific backbones; (2) a two-stage fusion scheme that enforces cross-modal consistency via contrastive alignment and then performs adaptive, task-aware fusion through a gating mechanism and a \emph{Q-bottleneck} that selectively compresses information for downstream reasoning; and (3) a lightweight, load-balanced Mixture-of-Experts at prediction time that enables efficient specialization to diverse semantic patterns. Extensive experiments demonstrate that \textsc{FLUID} attains \(91\%\) accuracy on the GLAMI-1M benchmark, significantly outperforming prior baselines and exhibiting strong resilience to label noise, long-tail class imbalance, and semantic heterogeneity. Targeted ablation studies corroborate both the individual and synergistic benefits of the proposed components, positioning \textsc{FLUID} as a scalable, noise-resilient solution for multimodal product classification.



\end{abstract}

\section{Introduction}

In today’s digital landscape, understanding multimodal data such as images and textual descriptions, is essential to applications like content moderation, recommendation systems, and media organization. Yet, classifying such data is inherently difficult, especially when inputs are inconsistent, or incomplete \cite{jain2021overview,baltruvsaitis2018multimodal}.

Traditional unimodal models, which process only a single modality—such as textual descriptions or visual content—often falter when essential contextual cues are absent from the input. For example, an image without accompanying text may lack semantic clarity, while text alone may miss crucial visual attributes \cite{hendricks2018grounding}. To mitigate this, multimodal models have emerged, aiming to integrate complementary signals across different data types. These models typically employ dedicated encoders for each modality, followed by a fusion module that aggregates features into a shared latent space for downstream tasks \cite{lu2019vilbert}.

While this framework promises richer and more informative representations, it also introduces new challenges. Specifically, many existing systems struggle with modality imbalance, where one modality dominates or overwhelms the other; weak cross-modal alignment, where features from different modalities are poorly synchronized; and sensitivity to noise, particularly when one of the modalities contains irrelevant or misleading signals \cite{rui2020research,kiela2018learning}. These limitations hinder the generalization ability of multimodal models, especially under real-world conditions characterized by incomplete, ambiguous, or noisy data \cite{hadsell2006dimensionality}.

To address these challenges, we propose FLUID—a robust multimodal classification architecture that strengthens both representation learning and modality interaction. FLUID introduces two Q-Transform blocks to extract condensed, high-salience features from each modality, followed by a gating mechanism and a Q-Bottleneck module to filter and route task-relevant signals. This design allows the model to prioritize useful information while suppressing noise, improving performance in real-world scenarios. In place of a standard feedforward classifier, FLUID incorporates a lightweight Mixture-of-Experts (MoE) layer, enabling adaptive expert selection for more flexible decision-making under diverse inputs.

Empirically, FLUID achieves strong results—outperforming prior baselines by at least 13\% on noisy, large-scale multimodal benchmarks—demonstrating its effectiveness in practical, high-variance settings.

\section{{Related works}}

Recent advances have highlighted the potential of multimodal learning for product categorization in e-commerce by leveraging both visual and textual data. Several works have explored this direction by incorporating enhanced attention mechanisms and cross-modal interactions.

Hung et al.~\cite{hung2024enhanced} proposed an attention-based multimodal framework that refines representations for each modality before cross-modal interaction. Their approach effectively mitigates noisy or inconsistent product descriptions by enabling the model to focus on informative regions of each input.

In the fashion domain, Han et al.~\cite{han2023fame} introduced FAME-ViL, a multi-task vision-language model that handles diverse downstream tasks using shared visual-linguistic representations and task-specific cross-attention layers. This architecture improves robustness and generalization in fashion-related classification tasks. In addition, Wei et al.~\cite{wei2020multi} presented the Multi-Modality Cross Attention Network (MMCA), which tackles image-sentence matching using fine-grained token- and region-level bidirectional attention. This design allows intricate semantic alignment, especially useful in distinguishing visually similar items with subtle textual differences. In another direction, Chen et al.~\cite{chen2023unified} proposed a unified vision-language framework using contrastive learning to project images, text, and their joint representations into a shared semantic space. 

Despite these advancements, current methods for classification task still face key limitations: (1) they often lack efficient mechanisms for transforming high-dimensional multimodal data into compact representations, making training computationally demanding; (2) fusion strategies tend to be naive or static, limiting the model’s ability to adapt to modality-specific signal strengths dynamically; and (3) many approaches fail to enforce alignment in a shared semantic space, reducing the synergy between modalities.

To address these gaps, we propose \textbf{\textsc{FLUID}}, a novel multimodal architecture designed to enhance representation quality and fusion efficiency. In summary, our main contributions are as follows:

\begin{itemize}
    \item \textbf{Enhancing Modality-specific Representation Technique:} We introduce Q–Transform—a novel approach, applying learnable query tokens to filter out irrelevant information and preserve the most salient features from image and text encoders.
    
    \item \textbf{Advancing Fusion Techniques:} We leverage contrastive learning to align the representation spaces of both modalities. Subsequently, a gating strategy and Q-Bottleneck module are employed to enable adaptive, flexible, and task-specific fusion, allowing the model to balance the contributions of each modality dynamically.

    \item \textbf{Specialized Output Prediction:} We adopt a lightweight Mixture-of-Experts (MoE) layer at the final stage. This enables expert specialization and conditional computation, improving output robustness and adaptability with minimal computational overhead.
    
    \item \textbf{Comprehensive Empirical Evaluation:} We extensively evaluate our method on the large-scale GLAMI-1M dataset, demonstrating its robust performance over existing approaches. In addition, thorough ablation studies confirm each proposed component's individual contribution and effectiveness.
\end{itemize}


\section{Motivation and Key ideas}
\begin{figure}
    \centering
    \includegraphics[width=0.55\textwidth]{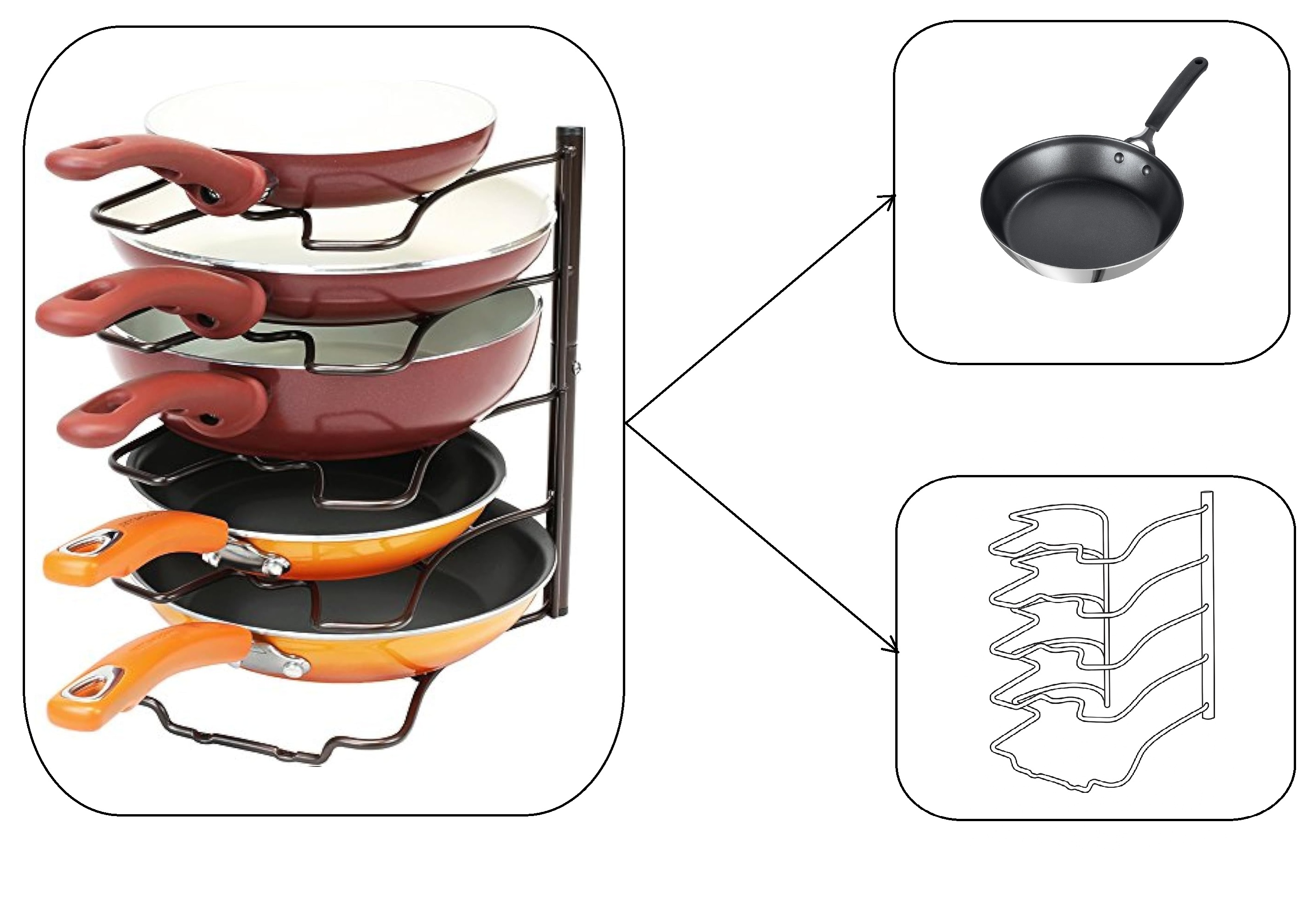} 
    \caption{A sample image on Amazon of a pan organizer rack which the unimodal models (rely solely on image information) may struggle to decide the label.
    }
    \label{pan}
\end{figure}
\subsection{Motivation}

\noindent\textit{Observation 1: Multimodal data are inherently ambiguous and prone to misinterpretation}

Classification tasks frequently depend on both visual and textual inputs; however, each modality individually often offers an incomplete, biased, or noisy representation~\cite{lu2019vilbert}. This limitation makes it difficult for unimodal systems to consistently produce accurate predictions, especially in complex, real-world settings, illustrated in Fig.\ref{pan}. This underscores the critical need for multimodal integration, where complementary signals across modalities are strategically fused to construct more robust, holistic representations. By leveraging each modality's strengths and mitigating weaknesses, multimodal approaches offer a more reliable foundation for downstream classification and decision-making.


\noindent\textit{Observation 2: Existing models struggle to preserve fine-grained cross-modal information}

While multimodal models have advanced significantly, most still rely on global representations—such as [CLS] tokens or pooling methods—to summarize each modality \cite{hung2024enhanced}\cite{gallo2020image}\cite{li2021align}. Though efficient, these techniques often overlook fine-grained token-level details that are crucial for nuanced understanding \cite{tsai2019multimodal}.

This creates a trade-off between compactness and semantic fidelity. Using all token embeddings could retain richer information, but at the cost of increased computational overhead and training instability, including issues like exploding gradients \cite{shazeer2017outrageously}.

Thus, a key challenge is developing selective mechanisms that can retain essential local information while preserving scalability and convergence. Solving this would enable more robust and semantically precise multimodal systems.

\noindent\textit{Observation 3: Limited modality synergy and adaptability in current fusion strategies}

While much effort has gone into improving unimodal encoders, fusion modules often remain simplistic—typically relying on direct concatenation of features \cite{hung2024enhanced}\cite{gallo2020image}\cite{zhang2021vinvl}. Recent work introduces cross-modal attention to enrich representations prior to fusion \cite{hung2024enhanced}\cite{tan2019lxmert}, yet most approaches still treat image and text as equally informative \cite{tsai2019multimodal}.

This symmetric treatment can hinder generalization, especially when one modality dominates in certain datasets.

To address this challenge, a more adaptive fusion mechanism is needed—one that can dynamically distill and reweight cross-modal cues based on context for improved flexibility and expressiveness.

\subsection{Key Ideas}

\noindent\textit{Key Idea 1: Compact and expressive modality representations via learnable queries (Q-Transform)}

To deal with the challenge mentioned in \textit{Observation 2}, we introduce Q-Transform—a set of $N$ learnable queries that extract the most salient features from each modality. This results in $N$ compact yet informative embeddings per modality, balancing efficiency and expressiveness.

\noindent\textit{Key Idea 2: Gated, modality-aware fusion in a unified embedding space}

Multimodal data often exhibit asymmetry, with one modality contributing more to the prediction. To handle this, we propose a gated fusion module that learns a gating vector to adaptively weigh each modality’s contribution. Prior to fusion, we align text and image embeddings into a shared latent space using contrastive learning, enabling coherent integration.

We then apply a lightweight Q-Bottleneck block on the fused tokens to refine task-relevant information. This design allows the model to dynamically adapt to input variability and focus on the most informative cross-modal cues.

\section{Methodology}
\subsection{Overview}

\begin{figure}
    \centering
    \includegraphics[width=1\textwidth,
    trim= 0.3cm 0.5cm 0.5cm 0.5cm,
    clip]
    {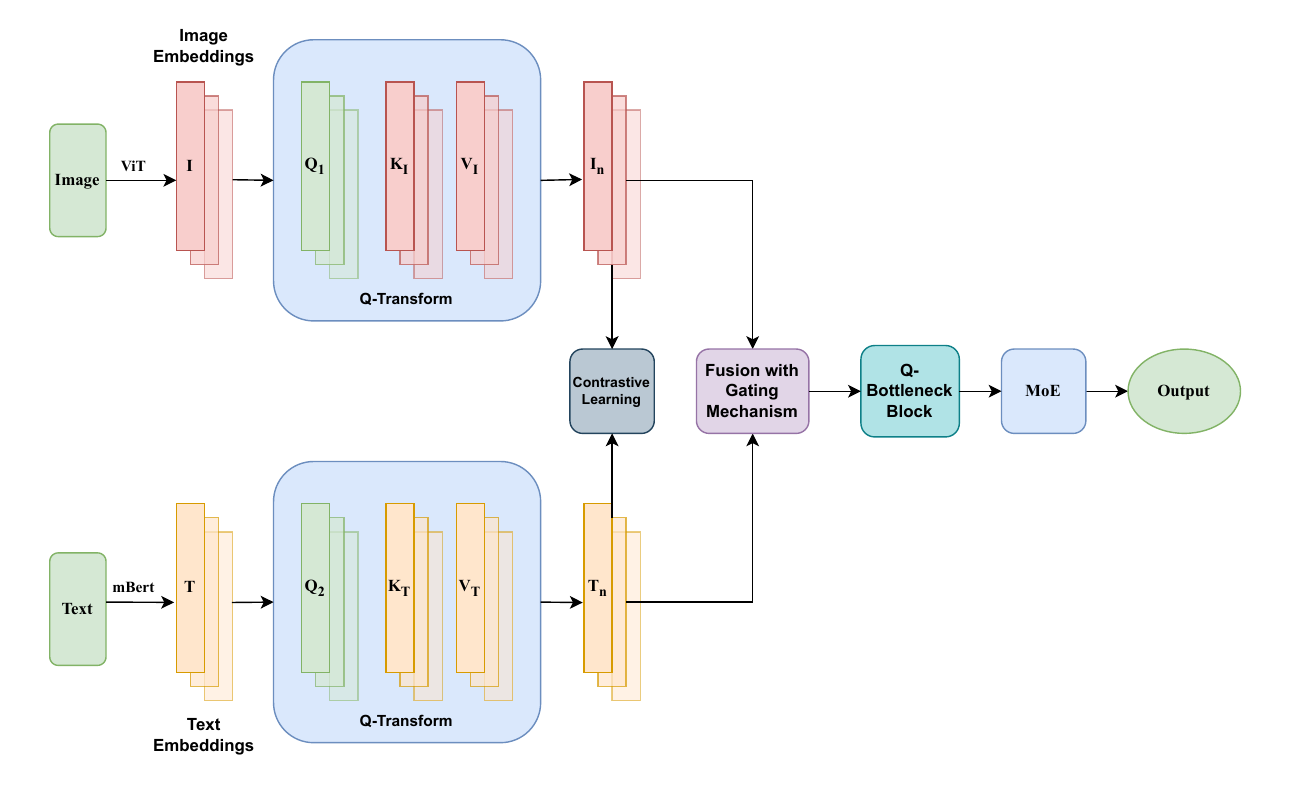}
    \captionsetup{font=normal}
    \caption{\textbf{Overview of the FLUID architecture.}
    Modality-specific encoders (ViT for images, mBERT for text) produce token sequences \(I\) and \(T\). Two sets of learnable queries \((Q_1, Q_2)\) compress these via multi-head attention into \(I_n\) and \(T_n\). Cross-modal embeddings are aligned contrastively, fused by a gated mechanism, and distilled through a Q-Bottleneck for MoE-based classification.}
    \label{model}
\end{figure}

As shown in Fig.~\ref{model}, FLUID begins by encoding each modality independently: mBERT for text and ViT for images, producing token embeddings $T$ and $I$. To extract compact, informative features, we apply the Q-Transform module, where fixed learnable queries $Q_1$ and $Q_2$ attend to $T$ and $I$, yielding compressed outputs $T_n$ and $I_n$. We incorporate a contrastive learning objective during training to encourage alignment between $I_n$ and $T_n$
in a shared embedding space, facilitating more coherent cross-modal fusion.The gating module fuses the two model outputs by computing adaptive weights that scale each contribution according to its task relevance. Finally, the fused representation is refined via a Q-Bottleneck module, where some learnable querys attends to the fused tokens, producing a task-specific embedding. This is then passed through a Mixture-of-Experts (MoE) layer to generate the final prediction.

\subsection{Encoder for image and text}
Our model takes as input an image with its corresponding textual description.

For the image encoder, we adopt ViT \cite{dosovitskiy2021imageworth16x16words}, a Transformer-based architecture widely used in image tasks. It captures global relationships across the image, producing an output matrix denoted as $I$.

\begin{equation}
    I = ViT(Image_{input}) \in \mathbb{R}^{197 \times 768}
\end{equation}

For the text encoder, we employ Multilingual BERT (mBERT) \cite{pires2019multilingualmultilingualbert} to address the multilingual nature of the dataset, which spans 13 languages \cite{Kosar_2022_BMVC}. Built upon BERT’s bidirectional transformer architecture \cite{devlin2019bertpretrainingdeepbidirectional}, mBERT provides language-agnostic contextual representations, producing a token embedding matrix \(T\).
\begin{equation}
    T = mBERT(Text_{input}) \in \mathbb{R}^{512 \times 768}
\end{equation}

\subsection{Q-transform technique}
Our model then incorporates two Q-Transform blocks, inspired by the Q-Former module introduced in BLIP-2\cite{li2023blip2}. This design has proven effective in extracting concise yet informative representations from input token sequences.

Previously, we denoted $I$ and $T$ as the token representations of the image and text modalities, respectively. In this stage, we apply a cross-attention mechanism, where $I$ and $T$ are linearly projected into key-value pairs: $(K_I, V_I)$ and $(K_T, V_T)$.

We then introduce two fixed, learnable query tokens, $Q_1$ and $Q_2$, which attend to the key-value matrices of their corresponding modalities. The operation is formally defined as follows:

\begin{equation}
    K_I = I W_I^K, V_I = I W_I^V, K_T = T W_T^K, V_T = T W_T^V
\end{equation}
\begin{equation}
    I_n = Attention(Q_1, K_I, V_I)
\end{equation}
\begin{equation}
    T_n = Attention(Q_2, K_T, V_T)
\end{equation}
where $W_I^K, W_I^V, W_T^K, W_T^V \in \mathbb{R}^{768 \times 768}$. $Q_1, Q_2 \in \mathbb{R}^{l \times 768}$. The attention function is indicated in \cite{vaswani2023attentionneed}, which formula is as follows:
\begin{equation}
    Attention(Q, K, V) = softmax(\frac{QK^T}{\sqrt{d}}) V
\end{equation}

The learnable query tokens can be interpreted as a set of fixed questions designed to extract the most task-relevant information from the input. The choice of the number of queries, denoted as $l$, is crucial. If $l$ is too small, the representation becomes overly compressed, similar to using a single [CLS] token; if $l$ is too large, it approximates using all tokens from each modality, reducing the effectiveness of filtering. Through empirical experimentation during training, we found that the optimal value for $l$ is 32, which results in modality-specific representations $I_n$ and $T_n$ that information balance and training efficiency.

\subsection{Fusion with Gating Mechanism}

At this stage, we obtain two modality-specific representation matrices: $I_n$ and $T_n$, each of dimension $l \times 768$, where $l = 32$ as determined in the previous step. However, not all token vectors contribute equally to the final prediction—a point we further analyze in the experiment \ref{uni-ablation}. To address this, we introduce a gating mechanism that adaptively fuses these representations into a unified feature space. This fusion takes into account the relative importance of each modality for a given sample, enabling richer and more balanced multimodal representations.

Specifically, we construct a gating vector $\mathbf{a} \in \mathbb{R}^l$, which is dynamically generated per sample, demonstrated in Fig.\ref{fusion}. The fused representation matrix $F$ is then computed as follows:

$$
F = \mathbf{a} \odot I_n + (1 - \mathbf{a}) \odot T_n
$$

where $\odot$ denotes scalar-wise multiplication. The computation of $\mathbf{a}$ involves two main steps:

First, for each input sample, we concatenate the representation matrices $I_n$ and $T_n$ along the feature dimension, resulting to a matrix of dimension $l\times(768\times2)$. This sample-specific combined representation is then passed through a lightweight feedforward neural network consisting of a single hidden layer. The hidden layer uses a ReLU activation to introduce non-linearity, and the output layer uses a sigmoid activation to ensure that the resulting gating vector $\mathbf{a} \in (0, 1)^l$. This gating vector is computed uniquely for each sample and dynamically adjusts the contribution of each token from the image and text modalities in the fusion process.

\subsection{Q-bottleneck block}
\begin{figure}
    \centering
    \includegraphics[width=0.7\textwidth]{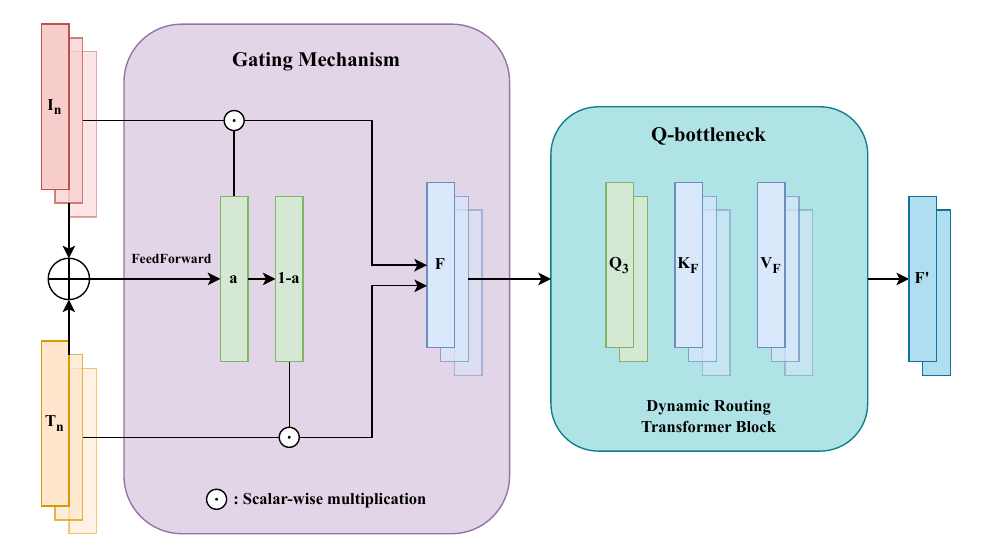}
    \captionsetup{font=normal}
    \caption{\textbf{The proposed Fusion module with advanced techniques:} Gating Mechanism and Q-bottleneck.
    }
    \label{fusion}
\end{figure}

At this stage, we obtain 32 high-quality vectors that encapsulate the most salient features from both text and image modalities. To effectively synthesize this information into a compact, task-relevant representation, we introduce the Q-Bottleneck module.
This block operates similarly to the Q-transform; however, the query matrix used in this module has a shape of $2 \times 768$. In this case, the input is the fused matrix $F$, and the output is a matrix $F'$ of dimension $2\times768$, illustrated in Fig.\ref{fusion}:
\begin{equation}
    K_F = F W_F^K, V_F = F W_F^V
\end{equation}
\begin{equation}
    F' = Attention(Q_3, K_F, V_F)
\end{equation}
where $W_F^K, W_F^V \in \mathbb{R}^{768 \times 768}$, $Q_3 \in \mathbb{R}^{2\times768}$. 

By using the query vectors $Q_3$ to attend over the matrix $F$, the model effectively evaluates the importance of each vector within $F$. As a result, the output vectors $F'$ retains only the most relevant information, reducing noise and enhancing the quality of predictions. Q-bottleneck and Q-transform function as a two-layer filtering mechanism that processes all input tokens and preserves the deepest, task-relevant information.

\subsection{Mixture-of-Experts for Output Prediction}

Instead of a conventional feedforward network, we employ a lightweight Mixture-of-Experts (MoE) layer to generate the final output probabilities. The MoE mechanism consists of multiple expert subnetworks, each trained to specialize in different data patterns. A gating network dynamically selects and weights a small subset of these experts for each input, enabling conditional computation and improved capacity without a proportional increase in computational cost.

Formally, given the fused representation $F'$, the output of the MoE layer is defined as:
\begin{equation}
N_{F'} = \sum_{i=1}^{top-k} G_i(F') \cdot E_i(F')
\end{equation}
where $E_i$ is the $i$-th expert, $G_i(F')$ is the gating weight assigned to $E_i$, and top-k is the number of selected experts. Each expert is a small MLP, and only a few are activated per input instance.

\subsection{Loss Function}

To train our multimodal classification model, we combine multiple loss components to guide both cross-modal alignment and final prediction.

We use the standard cross-entropy loss $\mathcal{L}_{CE}$ for classification, applied to the output of the fused representation branch.

We employ an image–text contrastive loss $\mathcal{L}_{\text{Contrast}}$ to pull matched embeddings together and repel mismatched ones, yielding tighter cross-modal alignment for subsequent fusion\cite{radford2021learningtransferablevisualmodels}.

Finally, to improve expert specialization and routing confidence in our MoE classifier, we add an auxiliary load balancing loss $\mathcal{L}_{\text{MoE}}$ that encourages uniform expert usage \cite{lepikhin2020gshardscalinggiantmodels}.
The total training objective is:
\begin{equation}
\mathcal{L}_{total} = \lambda_1 \mathcal{L}_{CE} + \lambda_2 \mathcal{L}_{\text{Contrast}} + \lambda_3 \mathcal{L}_{\text{MoE}}
\end{equation}
where $\lambda_1, \lambda_2, \lambda_3$ balance the contribution of each loss term. In our experiments, we set all weights equally to $1/3$.

\section{Empirical Evaluation}

To rigorously assess our proposed method, \textsc{FLUID}, we address three key research questions:

{\hangindent=2em \textbf{RQ1. [Effectiveness]} How does \textsc{FLUID} compare with current multimodal baselines in the various classification metrics in standard e-Commerce datasets?}

{\hangindent=2em \textbf{RQ2. [Multimodal Efficiency]} How effectively does \textsc{FLUID} improve performance compared to single-modality models (image-only and text-only)?}

{\hangindent=2em \textbf{RQ3. [Ablation analysis]} How critical do the components of our method contribute to its performance?}

All experiments were conducted on a system equipped with an NVIDIA A100 80GB PCIe GPU.

{\hangindent=2em \textbf{Dataset.} We evaluate our method on the \textbf{GLAMI-1M} multilingual large-scale image-text classification benchmark \cite{Kosar_2022_BMVC}, which consists of 1.11M fashion product records, each with an image, a description (in one of 13 languages), and a category label spanning 191 classes. This benchmark is challenging due to its real-world characteristics: extreme class imbalance (ranging from 81 to more than 75K samples per class), long-tailed label distribution, and noisy annotations ($25\%$ of training labels are machine generated)}.

{\hangindent=2em \textbf{Baselines.} We compare our proposed method, \textsc{FLUID}, against three strong baselines: (1) Improved-Attention~\cite{hung2024enhanced}, a recent enhancement of attention-based multimodal classification; (2) Early-Fusion~\cite{gallo2020image}, a representative baseline that concatenates modalities at the input level; and (3) a tailored adaptation of BLIP2 \cite{li2023blip2}, modified to align with the requirements of our classification task.}

{\hangindent=2em \textbf{Configure.} We set the number of experts in the MoE module to 16 and use top-k = 2. The model is trained for 15 epochs with a batch size of 128.}

{\hangindent=2em \textbf{Optimizer.} We use AdamW with a learning rate of $2 \times 10^{-5}$ and a weight decay of 0.01 to ensure stable updates and prevent overfitting.}

{\hangindent=2em \textbf{Evaluation Metrics.} We report standard classification metrics, including accuracy, precision, recall and F1-Score.}

\section{Effectiveness of \textsc{FLUID} (RQ1)}
We conducted a comparative evaluation on the GLAMI-1M dataset to assess the performance of the proposed \textsc{FLUID} model against several baselines, with detailed results provided in Table \ref{tab:rq1}. The evaluation unequivocally demonstrates that \textsc{FLUID} achieves substantially superior performance.

\begin{table}[H]
\small
\centering
\caption{Overall Performance Comparison Across All Metrics}\label{tab:rq1}
\begin{tabularx}{\columnwidth}{l *{4}{>{\centering\arraybackslash}X}}
\toprule
Method & Accuracy (\%) & Precision (\%) & Recall (\%) & F1-Score (\%) \\
\midrule
Early-Fusion~\cite{gallo2020image} & 63 & 53 & 40 & 41 \\
Improved-Attention~\cite{hung2024enhanced} & 66 & 56 & 42 & 45 \\
Modified BLIP2 \cite{li2023blip2} & 78 & 64 & 62 &63 \\

\textbf{FLUID} & \textbf{91} & \textbf{87} & \textbf{87} & \textbf{87} \\
\bottomrule
\end{tabularx}
\end{table}

In general, \textsc{FLUID} achieves a remarkable 91\% Accuracy, substantially outperforming Modified BLIP2 (78\%), Improved-Attention (66\%), and Early-Fusion (63\%). This superior performance is consistently reflected across other key metrics, with \textsc{FLUID} attaining 87\% Precision, 87\% Recall, and an F1-Score of 87\%. These results correspond to relative improvements of 23--47\% over existing methods, underscoring the effectiveness and robustness of our proposed approach.

This substantial performance gap is not incidental but a direct consequence of \textsc{FLUID}'s architectural innovations in representation extraction and fusion. By incorporating the Q-Transformer, our model focuses exclusively on the most informative tokens from each modality, effectively preserving salient content while discarding irrelevant noise. Additionally, integrating contrastive alignment and a gating mechanism enables \textsc{FLUID} to align image and text representations more effectively and adaptively weight their contributions—particularly crucial in the face of modality imbalance (which we further analyze in Experiment \ref{uni-ablation}). Finally, the lightweight Q-Bottleneck block enhances the fused representation by filtering out residual task-irrelevant information. This synergistic design empowers \textsc{FLUID} to overcome the limitations of conventional models and achieve superior, consistent results.

\renewcommand{\arraystretch}{1.2}

\section{Multimodal Efficiency (RQ2) }
\label{uni-ablation}
As discussed in our earlier observations, we conducted controlled experiments comparing the full \textsc{FLUID} model with its unimodal variants (text-only and image-only).
\setlength{\tabcolsep}{1pt}             

\begin{table}[H]
\small
\centering
\caption{Performance Comparison Between Multimodal and Unimodal Variants}
\label{tab:ablation-modality}
  \begin{tabularx}{\columnwidth}{@{}l *{4}{>{\centering\arraybackslash}X}@{}}
\toprule
Method & Accuracy (\%) & Precision (\%) & Recall (\%) & F1-Score (\%) \\
\midrule
Image-only & 46 & 28 & 22 & 21 \\
Text-only & 73 & 63 & 54 & 55 \\
\textbf{FLUID} & \textbf{91} & \textbf{87} & \textbf{87} & \textbf{87} \\
\bottomrule
\end{tabularx}
\end{table}

Table~\ref{tab:ablation-modality} confirms the superiority of \textsc{FLUID} over unimodal variants. While the F1-score of both image-only and text-only models drops sharply—especially under the noisy, long-tailed distribution of GLAMI-1M—\textsc{FLUID} maintains strong and stable performance, achieving up to 66\% and 32\% relative improvements, respectively. These results highlight the effectiveness of multimodal fusion and the importance of explicitly modeling cross-modal interactions. Furthermore, the text-only model consistently outperforms the image-only variant, revealing an imbalance in unimodal contributions and motivating the need for a dynamic fusion strategy.

\section{Ablation Study (RQ3) }

\textsc{FLUID} comprises five components: (1) Contrastive loss, (2) Q-transform, (3) Gating mechanism, (4) Q-bottleneck, and (5) MoE. In this experiment, we investigate each component's contribution by systematically replacing each of them. The figures will be provided in Table \ref{tab:ablation-FLUID}.

\begin{table}[H]
\small
\centering
\caption{Ablation Study of \textsc{FLUID}: Overall Results Across All Metrics}
\label{tab:ablation-FLUID}
\begin{tabularx}{\columnwidth}{l *{4}{>{\centering\arraybackslash}X}}
\toprule
Method & Accuracy (\%) & Precision (\%) & Recall (\%) & F1-Score (\%) \\
\midrule
Without Contrastive & 75 \textcolor{red}{{(-16\%)}} & 68 \textcolor{red}{{(-19\%)}} & 66 \textcolor{red}{{(-21\%)}} & 67 \textcolor{red}{{(-20\%)}}\\
Without Q-Transform & 87 \textcolor{red}{{(-4\%)}} & 83 \textcolor{red}{{(-4\%)}}& 80 \textcolor{red}{{(-7\%)}} & 81 \textcolor{red}{{(-6\%)}} \\
Without Gating & 88 \textcolor{red}{{(-3\%)}} & 85 \textcolor{red}{{(-2\%)}} & 84 \textcolor{red}{{(-3\%)}} & 84 \textcolor{red}{{(-3\%)}} \\
Without Q-Bottleneck & 75 \textcolor{red}{{(-16\%)}} & 67 \textcolor{red}{{(-20\%)}} & 63 \textcolor{red}{{(-24\%)}} & 64 \textcolor{red}{{(-23\%)}} \\
Without MoE & 88 \textcolor{red}{{(-3\%)}} & 85 \textcolor{red}{{(-2\%)}} & 84 \textcolor{red}{{(-3\%)}} & 85 \textcolor{red}{{(-2\%)}} \\
\textbf{FLUID} & \textbf{91} & \textbf{87} & \textbf{87} & \textbf{87} \\
\bottomrule
\end{tabularx}
\end{table}

\textbf{Contrastive loss:} Overall, performance drops markedly when this mechanism is ablated: Accuracy 91\%\,$\rightarrow$\,75\%, Precision 87\%\,$\rightarrow$\,68\%, Recall 87\%\,$\rightarrow$\,66\%, and F1‑Score 87\%\,$\rightarrow$\,67\%. 
This demonstrates that contrastive alignment anchors text and image embeddings onto a shared manifold, providing cleaner, better aligned input to the fusion layers at a negligible cost.

\textbf{Q-transform:} Table~\ref{tab:ablation-FLUID} shows that removing the Q-transform module leads to notable performance drops: Accuracy falls to 87\%, Precision to 83\%, Recall to 80\%, and F1‑Score to 81\%. These results highlight its importance in selecting salient, task-relevant tokens. Without it, the model relies only on [CLS] tokens, which may overlook fine-grained semantic cues. The Q-transform enhances fusion quality by focusing on informative features and filtering out noise.

\textbf{Gating Mechanism:} This component brings consistent gains, +3\% in Accuracy, F1-Score, and Recall, and +2\% in Precision. Though seemingly modest, these improvements translate to thousands of correct predictions on GLAMI‑1M. The results confirm that adaptive gating helps prioritize the most informative modality, enhancing overall decision quality.

\textbf{Q-Bottleneck:} Removing this module leads to the sharpest performance drop—Accuracy falls to 75\%, Precision to 67\%, Recall to 63\%, and F1‑Score to 64\%—highlighting its critical role. By compressing purified features into a compact subspace, it reduces cross-modal noise and tightens class boundaries. Without it, the model struggles with a noisy representation space, severely degrading all metrics.

\textbf{MoE Layer:} Removing MoE Layer results in consistent performance drops across all metrics: Accuracy decreases by 3\%, Precision by 2\%, Recall by 3\%, and F1-Score by 2\%. By leveraging conditional computation and expert specialization, the MoE module allows the model to better adapt to diverse patterns in multimodal data without incurring significant computational overhead.

\textbf{Summary.} The ablation study underscores the critical importance of each component within the \textsc{FLUID} architecture. These modules are not isolated; their effectiveness emerges through synergistic interaction. For instance, the presence of the Q-transform enhances the gating mechanism's ability to modulate information flow, enabling the Q-bottleneck to distill the most informative features more effectively and deliver them to specialized experts. This interdependence among components ultimately drives the superior performance of \textsc{FLUID}.

\section{Conclusion and Future works}
In this study, we proposed a novel model named \textsc{FLUID}, designed to enhance multimodal classification performance. First, we effectively extract visual and textual information from inputs by using base encoders followed by Q-transform blocks. Then, we employ a contrastive loss to align the embedding spaces of the two modalities. Finally, we utilize an advanced fusion module that can dynamically select the most task-relevant features thanks to the gating mechanism and the Q-bottleneck block. To further increase model capacity and flexibility without incurring heavy computation, we incorporate a lightweight Mixture-of-Experts (MoE) layer, allowing expert specialization and conditional computation during final prediction. Experiments on the GLAMI-1M dataset demonstrate that \textsc{FLUID} significantly outperforms existing state-of-the-art methods, achieving superior accuracy, precision, recall, and F1 score, while exhibiting strong generalization to noisy and diverse real-world data. 

For future work, we plan to explore multi-stage hierarchical fusion, where fusion occurs at multiple semantic levels (e.g., object, attribute, and global context), enabling the model to better capture both fine-grained and high-level cross-modal dependencies.

\printbibliography

\end{document}